\documentclass[aps,prb,twocolumn,superscriptaddress,showpacs]{revtex4}
\usepackage{graphicx}
\begin{document}
\title{\bf Character of magnetic excitations in a quasi-one-dimensional
antiferromagnet near the quantum critical points: Impact on
magneto-acoustic properties}
\author{O.~Chiatti}
\author{A.~Sytcheva}
\author{J.~Wosnitza}
\author{S.~Zherlitsyn}
\affiliation{Hochfeld-Magnetlabor Dresden (HLD), Forschungszentrum
Dresden-Rossendorf, D-01314 Dresden, Germany}
\author{A.~A.~Zvyagin}
\affiliation{Max-Planck-Institut f\"ur Physik komplexer Systeme, D-01187
Dresden, Germany}
\affiliation{B.~Verkin Institute for Low Temperature Physics and
Engineering of the NAS of Ukraine, Kharkov, 61103, Ukraine}
\author{V.~S.~Zapf}
\author{M.~Jaime}
\affiliation{National High Magnetic Field Laboratory, Los Alamos
National Laboratory, Los Alamos, NM 87545, USA}
\author{A. Paduan-Filho}
\affiliation{Instituto de Fisica, Universidade de S$\tilde{a}$o
Paulo, 05315-970 S$\tilde{a}$o Paulo, Brazil}

\date{\today}

\begin{abstract}
We report results of magneto-acoustic studies in the quantum
spin-chain magnet NiCl$_2$-4SC(NH$_2$)$_2$ (DTN) having a
field-induced ordered antiferromagnetic (AF) phase. In the vicinity
of the quantum critical points (QCPs) the acoustic $c_{33}$ mode
manifests a pronounced softening accompanied by energy dissipation
of the sound wave. The acoustic anomalies are traced up to $T >
T_N$, where the thermodynamic properties are determined by fermionic
magnetic excitations, the ``hallmark'' of one-dimensional (1D) spin
chains. On the other hand, as established in earlier studies, the AF
phase in DTN is governed by bosonic magnetic excitations. Our
results suggest the presence of a crossover from a 1D fermionic to a
3D bosonic character of the magnetic excitations in DTN in the
vicinity of the QCPs.
\end{abstract}

\maketitle

The interest in quasi-1D quantum spin systems has grown considerably
during the last decade. This is fostered by the progress in
preparing materials with well-defined 1D spin subsystems and the
possibility of analyzing the experimental data with the help of
non-perturbative theories for 1D models. \cite{Zb} In addition, such
systems often manifest quantum phase transitions at $T$=0 which are
governed by parameters other than the temperature. True 1D models do
not exhibit any long-range order at finite temperatures. \cite{Zb}
Real quasi-1D antiferromagnetic (AF) materials, containing weakly
coupled spin chains with gapless spectra of their low-lying
excitations, are usually magnetically ordered at low temperatures.
At temperatures higher than the Ne\'el temperature, $T_N$, but of
the order of the exchange constant, these systems behave as quantum
spin chains, where any long-range magnetic order is destroyed by
enhanced quantum fluctuations. \cite{Zb} One should note that
quasi-1D magnets, in which the low-energy eigenstates of their 1D
subsystems have spin gaps, usually do not manifest long-range
magnetic ordering. \cite{Reg} However, an external magnetic field
can close the spin gap, $\Delta$, and for $H > H_c \sim \Delta$ a
quantum phase transition to a phase with gapless spin excitations
takes place. A further increase of the field yields a second quantum
phase transition to a spin-polarized phase at $H > H_s$. In the
spin-polarized phase the low-energy excitations are also gapped.
Hence, the magnetically ordered phase can be observed in the field
domain where spin excitations are gapless, and the N\'eel
temperature in such systems is field dependent. The magnetic
susceptibilities of a quasi-1D spin system in mean-field
approximation can be written as
\begin{equation}
\chi_{\bf q}^{\alpha} = {(\chi_{\bf q}^{\alpha})^{(1)}(T)\over 1 -
ZJ_{\perp}({\bf q}) (\chi_{\bf q}^{\alpha})^{(1)}(T)}  \ ,  \
\label{chi3d}
\end{equation}
where the superscript $(1)$ denotes the susceptibility of one
chain, $\alpha =x,y,z$, $J_{\perp}$ is the weak interchain
exchange constant, $Z$ is the coordination number, and {\bf q} is
the wave vector. The quasi-1D spin system becomes ordered when the
denominator becomes zero (which defines $T_N(H)$).

The low-$T$ thermodynamics of a state with long-range magnetic order
is determined by bosonic excitations, magnons. Recently, several
groups have observed phenomena in some AF systems that have been
interpreted as Bose-Einstein condensation (BEC) of magnons, viz., as
a thermodynamically large number of magnons in the same ground
state. \cite{Seb, Rad, Rue} For quasi-1D spin systems at $T > T_N$
the projection of a single spin may have only a limited number of
values (e.g., two values for spin-1/2 systems, three values for
spin-1, etc.). That is why thermodynamic properties of, e.g., AF
spin-1/2 chains are often determined by low-energy eigenstates which
behave as interacting {\em fermions}. \cite{Zb} The fermionic nature
of the low-energy excitations of these quantum spin chains with only
short-range correlations is related to the limited number of
projections of each spin. Very recently it has been shown \cite{MHO}
that the behavior of a spin-1 spin-gapped system in magnetic fields
close to field-induced quantum critical points (QCP's) can be
described by free fermions as well. The fermionic nature of these
excitations is also related to the limited number of projection
values at each spin-1 site. The fermionic behavior of excitations is
characteristic for quantum spin chains with short-range
correlations. From this perspective it is very interesting to study
the behavior of a spin-gapped quasi-1D AF system in the vicinity of
$T_N$, close to $H_c$ and $H_s$. Here, low-lying excitations of the
quasi-1D system should change their statistical properties from
fermionic, at $T > T_N(H)$, to bosonic, at $T < T_N(H)$. Hence, by
varying $H$ and $T$ one may observe features in the same spin system
characteristic either to fermions or bosons.

One of the best candidates for studying such a crossover in the
excitation statistics is the spin-1 system dichloro-tetrakis
thiourea-nickel(II), NiCl$_2$-4SC(NH$_2$)$_2$, known as DTN.
Recently, some features in the magnetically ordered phase of DTN
at $H \ge H_c$ and $T <T_N(H)$ were interpreted as BEC of spin
degrees of freedom. \cite{Zapf} The bosonic character of the spin
excitations in DTN in the magnetically ordered phase has been
corroborated and is considered to be a well established fact. In
this work, we study magnetic and magneto-acoustic characteristics
of DTN near the critical values of $H_c$ and $H_s$. We show that
the behavior of the observed properties outside of the AF phase
can be well described by an effective {\em fermionic} model of
low-lying spin excitations. In this way DTN manifests 1D fermionic
character of spin excitations at $T > T_N(H)$. This fact, together
with previous results, showing bosonic 3D behavior of magnetic
excitations in DTN for $T < T_N(H)$, \cite{Zapf} leads us to the
conclusion that a crossover from fermionic to bosonic features of
the low-lying magnetic excitations takes place at $T_N$ near the
quantum critical points.

Following Ref.~[\onlinecite{MHO}] we describe the spin-1 chain at
low excitation densities using an effective free-fermion theory with
two branches of low-energy states. Two branches are used because the
strong single-ion ``easy-plane'' magnetic anisotropy $D$ observed in
DTN \cite{Zv} splits the spin triplet of the spin-gap modes, and
makes one of them ineffective at the critical fields. \cite{MHO} The
two fermionic branches have features at $H_c$ and $H_s$, while for
$D \ll T$ the contribution of the third branch is exponentially
small and can be neglected in our approximation. Both critical
fields are related only to the lowest branch of our model (they
correspond to van Hove singularities, connected with two edges of
that branch). However, the field dependence is present also in the
temperature-dependent factors of both branches.
\begin{figure}
\begin{center}
\includegraphics[height=5.0cm,width=0.45\textwidth]{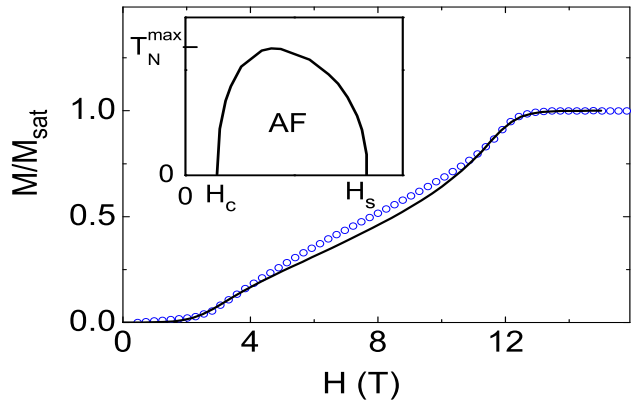}
\end{center}
\caption{(Color online) Low-temperature magnetization ($T = 0.6$~K)
of DTN as a function of external magnetic field $H
\parallel$[001] (circles). \cite{Arm} The line is the result of the
free-fermion effective theory. The inset sketches the temperature
- magnetic field phase diagram of DTN. \cite{Zapf} The quantum
critical points are $H_c \approx 2.1$~T and $H_s \approx 12.6$~T.
The maximum temperature of the AF order is $T_N^{max} \approx
1.2$~K.} \label{fig1}
\end{figure}
In Fig.~1 the solid line shows the calculated field dependence of
the magnetization of the quasi-1D spin system (1D subsystems are
considered within this effective free-fermion model) for $T$
slightly above the phase boundary $T_N(H)$ (see inset in Fig.~1),
where the susceptibility of the quasi-1D system diverges. For
comparison we also plot experimental data taken at $T$ = 0.6~K.
\cite{Arm} Note that $T_N(H) <$ 0.6~K in the vicinity of the quantum
critical points $H_c$ and $H_s$, since there is a line of phase
transitions $T_N(H)$ with $T_N(H_c)=T_N(H_s)=0$ and the system in
our model is {\em not} in the magnetically ordered phase inside of
the interval $H^\prime_c \le H \le H^\prime_s$. $H^\prime_c$ and
$H^\prime_s$ are the critical fields at non-zero $T$. There is a
good agreement between the effective free-fermion theory and the
experimental data at $H < H^\prime_c$, $H > H^\prime_s$, and near
the critical values of $H$. On the other hand, inside of the
interval $H^\prime_c < H < H^\prime_s$, the real system is ordered,
$T_N(H)>$ 0.6~K, and our 1D fermionic description cannot be applied.

Ultrasonic investigations are a powerful experimental technique to
study various phase transitions and critical phenomena. This
technique is well established as an important tool for the
investigation of low-dimensional spin systems. \cite{L} Spin-lattice
interactions are responsible for the attenuation of acoustic waves
and influence the sound velocity in magnetic crystals. These
interactions are connected either with a strain modulation of the
exchange interactions or with a magnetostrictive coupling of a
single-ion type. \cite{L} We have performed measurements of the
relative change of the sound velocity and attenuation in DTN, using
a phase-sensitive detection technique based on a standard pulse-echo
method with a set-up similar to the one described in
Ref.~[\onlinecite{L}]. DTN has a tetragonal crystallographic
symmetry (space group I4) with two formula units in the unit cell.
The investigated single crystal has a size of about $2\times 2\times
4.1$~mm$^3$. Since the as-grown surfaces of the crystal were smooth
and parallel, we glued piezoelectric film transducers directly to
the surfaces normal to the crystallographic [001] direction, without
any additional sample polishing. This geometry corresponds to the
longitudinal acoustic $c_{33}$ mode, with propagation direction and
polarization along the spin chains. A number of ultrasonic echoes
have been detected. The absolute value of the sound velocity at
liquid-helium temperature has been determined as $v_l = 2640\pm
20$~m/s. Note that the measurement accuracy for a relative change of
sound velocity is of the order of 10$^{-6}$. The sample-length
change is relatively small for the applied temperatures and magnetic
fields. \cite{Za1} Therefore, we did not have to take into account
any length-change corrections to the sound velocity. The data have
been collected using the ultrasonic signal at 78~MHz. The magnetic
field was applied along the [001] direction, i.e., parallel to the
sound-propagation direction.
\begin{figure}
\begin{center}
\includegraphics[width=0.45\textwidth]{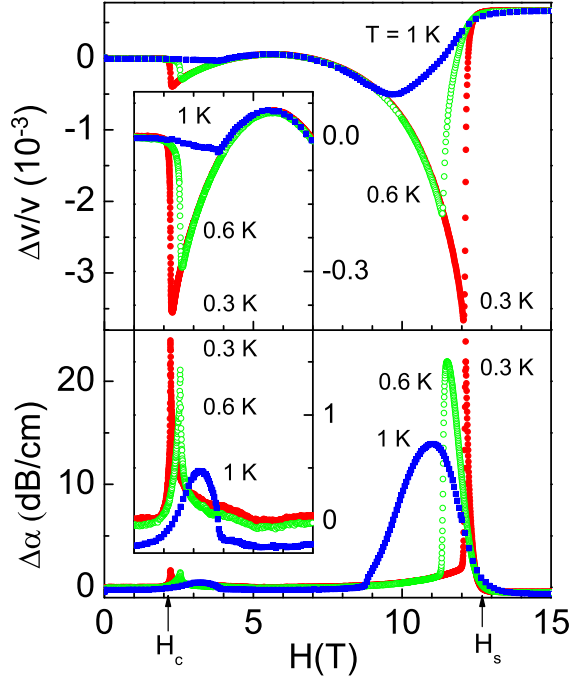}
\end{center}
\caption{(Color online) Field dependence of the relative change of
the sound velocity (top) and of the sound attenuation (bottom) of
the acoustic $c_{33}$ mode in DTN at $T$ below $T_N^{max}$. The
magnetic field was applied along the [001] axis. The ultrasonic
frequency was 78 MHz. The insets show the sound velocity and
attenuation in the vicinity of $H_c$ in enlarged scale.}
\label{fig2}
\end{figure}
Figure 2 shows the magnetic-field dependence of the relative change
of the sound velocity and attenuation of the $c_{33}$ mode in DTN
for $T$ below the maximum of $T_N^{max} \approx 1.2$~K. There is a
pronounced softening of the $c_{33}$ mode in the vicinity of both
critical fields, though the anomaly at $H^\prime_s$ is approximately
one order of magnitude larger than that at $H^\prime_c$. There is a
relative increase of $\Delta v/v$ = 7$\times$10$^{-4}$ between the
sound velocity at $H=0$ and $H
> 12.6$~T, where all spins are polarized. The relative decrease of the
sound velocity reaches about 4$\times$10$^{-3}$ at 12~T and 0.3~K.
The softening of the $c_{33}$ mode is accompanied by a peak in the
sound attenuation. Both the sound-velocity and sound-attenuation
anomalies become smaller and broader with increasing $T$. The $H$
dependence of the sound velocity in the ordered phase (far from the
critical regions) resembles $c$-axis magnetostriction data.
\cite{Za1} However, the change in the sound velocity cannot be
explained by the lattice-parameter change, since the length change
observed in Ref.~\onlinecite{Za1} is too small.
\begin{figure}
\begin{center}
\includegraphics[width=0.45\textwidth]{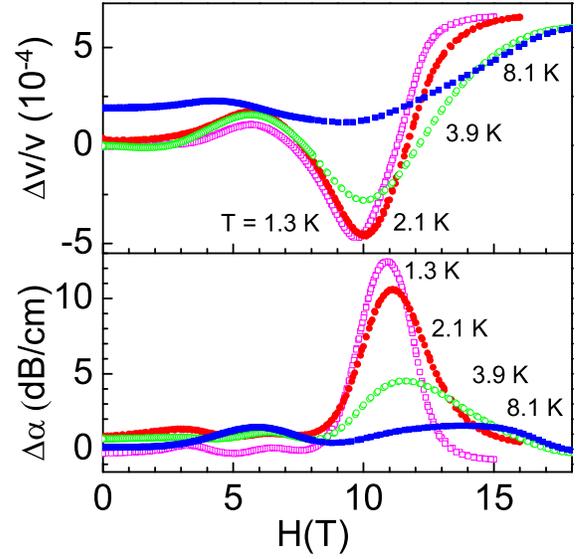}
\end{center}
\caption{(Color online) Field dependence of the relative change of
the sound velocity (top) and of the sound attenuation (bottom) of
the acoustic $c_{33}$ mode in DTN for $T$ above $T_N^{max}$.}
\label{fig3}
\end{figure}
In Fig.~3, we show the field dependence of the sound velocity and
attenuation of the $c_{33}$ mode in DTN at various temperatures
above $T_N^{max}$. One can see some transformation of the acoustic
anomalies by moving from $T < T_N^{max}$ to $T > T_N^{max}$. Here,
the softening of the $c_{33}$ mode disappears in the vicinity of
$H_c$; only a smooth increase in the sound velocity is detected.
Close to $H_s$ one can still observe a minimum in the sound velocity
and a maximum in the sound attenuation, but those anomalies are
smaller in amplitude and broader than the corresponding ones
measured below $T_N^{max}(H)$ (Fig.~2).

In magnetic materials the dominant contribution to the
spin-lattice interactions mostly arises from the
exchange-striction coupling. In our calculations we assumed that
in DTN the spatial dependence of the magnetic anisotropy constant
is weaker than the spatial dependence of the exchange integrals.
In this case, one can expect that only longitudinal sound waves
interact with the spin subsystem.
\begin{figure}[h]
\begin{center}
\includegraphics[width=0.4\textwidth, height=0.5\textwidth]{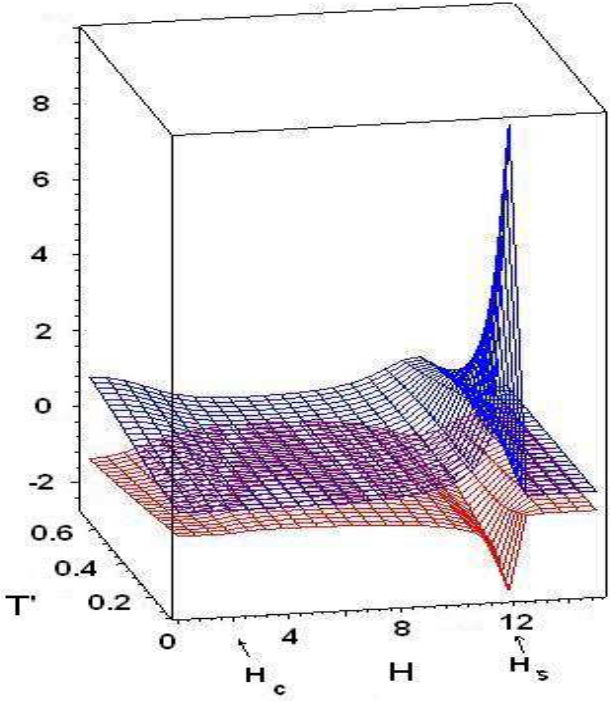}
\end{center}
\caption{(Color online) Attenuation (upper surface, blue) and
relative change of the velocity (lower surface, red) of the
longitudinal sound versus $H$ and $T^\prime$, calculated in the
framework of the proposed theory (arbitrary units were used for all
parameters, see text for details).} \label{fig4}
\end{figure}
According to Ref.~\onlinecite{TM}, the relative renormalization of
the longitudinal sound velocity can be written as $(\Delta v/v) =
- (A_1 +A_2)/ (N\omega_{\bf k})^2$, where
\begin{eqnarray}
&&A_1 = 2 |G_0^z({\bf k})|^2 \langle S_0^z \rangle^2 \chi_0^z + T
\sum_{\bf q} \sum_{\alpha =x,y,z} |G_{\bf q}^{\alpha} ({\bf k})|^2
(\chi_{\bf q}^{\alpha})^2 \ ,
\nonumber \\
&&A_2 = H_0^z({\bf k}) \langle S_0^z \rangle^2 +{T\over
2}\sum_{\bf q} \sum_{\alpha =x,y,z} H_{\bf q}^{\alpha} ({\bf k})
\chi_{\bf q}^{\alpha} \ . \label{otn}
\end{eqnarray}
Here, $N$ is the number of spins in the system, $\omega_{\bf k} =v
k$ is the low-$k$ dispersion relation with sound velocity $v$ in the
absence of spin-phonon interactions, $\langle S^z_0 \rangle$ is the
average magnetization along the direction of the magnetic field,
$\chi_{\bf q}^{x,y,z}$ are non-uniform magnetic susceptibilities,
and the subscript $0$ corresponds to $q=0$. In the framework of our
effective free-fermion model the temperature and magnetic-field
dependence of the uniform susceptibility of one spin chain can be
written as
\begin{eqnarray}
&&(\chi_0^z)^{(1)} = {8\over \pi T}\int_{H_c}^{H_s} {x dx\over
\sqrt{(H_s^2-x^2)(x^2-H_c^2)}} \times
\nonumber \\
&&{1 + \cosh(H/T)\cosh(x/T)\over [\cosh (H/T) +\cosh(x/T)]^2} \ ,
\label{1d}
\end{eqnarray}
where we set the units for the effective $g$-factor, Bohr's
magneton, and Boltzmann's constant equal to 1. For spin systems with
AF interactions the main contribution to the summation over ${\bf
q}$ in Eqs.~(2) comes from terms with $q =\pi$,
\begin{eqnarray}
&&(\chi_{\pi}^z)^{(1)} = {8\over \pi}\int_{H_c}^{H_s} {dx\over
\sqrt{(H_s^2-x^2)(x^2-H_c^2)}}\times
\nonumber \\
&&{\sinh (x/T)\over \cosh (H/T) +\cosh(x/T)} \ . \label{1dPi}
\end{eqnarray}
To calculate magnetic susceptibilities of the quasi-1D spin system
we use Eqs.~(1), (3) and (4).

The renormalization is proportional to the spin-phonon coupling
constants
\begin{eqnarray}
&&G_{\bf q}^{\alpha} = {1\over m}\sum_n e^{i{\bf q}{\bf
R}_{nm}}\left(e^{i{\bf k}{\bf R}_{nm}} -1\right) {\bf e}_{\bf k}
{\partial J_{mn}^{\alpha}\over
\partial {\bf R}_{m}} \ ,
\nonumber \\
&&H_{\bf q}^{\alpha} = {1\over m}\sum_n e^{-i{\bf q}{\bf
R}_{nm}}\left(e^{i{\bf k}{\bf R}_{nm}} -1\right)\left(e^{-i{\bf
k}{\bf R}_{nm}} -1\right) \times
\nonumber \\
&&{\bf e}_{\bf k}{\bf e}_{\bf -k}{\partial^2 J_{mn}^{\alpha}\over
\partial {\bf R}_{n}\partial {\bf R}_m} \ . \label{sphon}
\end{eqnarray}
Here, $m$ is the mass of the magnetic ion, $J_{mn}^{\alpha}$ denote
(anisotropic, generally speaking) exchange integrals, ${\bf e}_{\bf
k}$ is the polarization of the phonon with wave vector ${\bf k}$,
and ${\bf R}_n$ is the position vector of the $n$-th site. \cite{TM}
Figure~4 (lower surface) shows the $H$ and $T'$ dependence of the
relative change of the longitudinal sound velocity of a quasi-1D
spin system calculated in the framework of the effective
free-fermion model. We fixed $H_c$ and $H_s$ and used arbitrary
units for $T^\prime$ in Fig.~4 (they are not equal to the
temperatures in the experiment). It is challenging to calculate
$\chi_{\bf q}^{x,y}$ in the framework of the used model. Clearly
they have to be smooth functions of $H$ and $T$, except at the line
$T_N(H)$. $G_{\bf q}^{\alpha}({\bf k})$ and $H_{\bf q}^{\alpha}({\bf
k})$ in Eqs.~(2) are also unknown for any $\alpha$ (one of the
coupling constants can be estimated using Ref.~\onlinecite{Za1}).
That is why, in order to obtain the results presented in Fig.~4, we
used $J_{\perp}(q)=0.18$ from Ref.~\onlinecite{Zv}, $\chi_{\bf
q=0}^{z}$ and $\chi_{\bf q=\pi}^{z}$ multiplied by some (not known)
values of the spin-phonon coupling constants, and $H^z(\bf k)$ two
times smaller than $G^z(\bf k)$. The temperature of the divergence
in the magnetic susceptibility of the quasi-1D system is generally
determined by anisotropic couplings between the spin chains (these
couplings are unknown). The divergence at $T_N(H)$, which we used in
our theory, does not depend on the direction of the order parameter.
Such divergences are present in a quasi-1D model, when any component
of the magnetic susceptibility (but with different phase-transition
temperatures, $T_N(H)$) is considered. Even in this approximation
our simplified theory reproduces the main features of the
experimentally observed behavior. Our model reproduces the
pronounced minimum at $H^\prime_s$, the almost field-independent
behavior at $H > H^\prime_s$ and $H < H^\prime_c$, the larger value
of $\Delta v/v$ for $H > H^\prime_s$ as compared to $H <
H^\prime_c$, and the maximum (with $\Delta v/v > 0$) in the interval
between $H^\prime_c$ and $H^\prime_s$. With increasing $T'$ the
features near the critical fields become weaker, the same way as it
was observed in the experiment (cf. Fig.~2 and Fig.~3). At the phase
boundary $T_N(H)$ the susceptibility of the quasi-1D system diverges
(see above), and our theory predicts very narrow and large peaks at
the critical values of $H$ (not shown). Therefore, for the sake of
clarity, the curves in Fig.~4 are not plotted starting from $T'=0$.
$T_N^{max}(H)$ in our units is $T'=0.02$. Concerning the other
values of $\chi_{\bf q}$ (i.e., ${\bf q} \ne 0, \pi$) we affirm, as
it was discussed above, that their inclusion does not affect the
qualitative behaviour of the sound velocity and attenuation.
Following Ref.~\onlinecite{TM} we also calculated the attenuation
coefficient for DTN,
\begin{eqnarray}
&&\Delta \alpha (\equiv \Delta \alpha_k) ={1\over Nv} \bigg[2
|G_0^z({\bf k})|^2 \langle S_0^z \rangle^2 \chi_0^z
{\gamma_0^z\over (\gamma_{0}^z)^2 + \omega_{\bf k}^2}
\nonumber \\
&&+ T \sum_{\bf q} \sum_{\alpha =x,y,z} |G_{\bf q}^{\alpha} ({\bf
k})|^2 (\chi_{\bf q}^{\alpha})^2{2\gamma_{\bf q}^{\alpha}\over
(2\gamma_{\bf q}^{\alpha})^2 + \omega_{\bf k}^2} \bigg] \ , \
\label{alpha}
\end{eqnarray}
where $\gamma_{\bf q}^{\alpha}$ are the relaxation rates, which can
be approximated by $\gamma_{\bf q}^{\alpha} =B/T\chi_{\bf
q}^{\alpha}$, where $B$ is a material-dependent constant (see
Ref.~\onlinecite{TM}). In our calculations we used the
approximation, in which the relaxation rates do not depend on the
direction and on the wave vector. The results are also presented in
Fig.~4 (upper surface). Here, our theory reproduces also the main
features observed in the experimental data: an abrupt increase of
the sound attenuation near the saturation field $H_s$, and damping
with increasing $T'$. All these findings demonstrate the important
role the fermionic magnetic excitations play in the vicinity of the
QCPs in DTN.

We also tried to reproduce the observed experimental results using
the scaling-like procedure, proposed in Ref.~\onlinecite{Hon}. In
the framework of that approach we can use
$(\chi^{x,y}_{\pi})^{(1)}$, which seems more accurate than the use
of $(\chi^{z}_{\pi})^{(1)}$ only. However, the agreement between the
theory and experiment was worse than for our effective free-fermion
model.

Generally speaking, one could as well use a bosonic, say the
Holstein-Primakoff, representation of spin operators \cite{HP} for
any temperatures in DTN. However, to describe the behavior of spins
for $T > T_N(H)$ one has to take into account all interactions
between these bosons (because the interactions are of the same
magnitude as the energy of the free bosons), which is impossible so
far. We do not know any other theory, bosonic or fermionic, which
can describe the behavior of the magnetization, sound velocity, and
attenuation in DTN better than the theory presented here. The
situation, e.g., in weakly coupled spin ladders, is very different
from the one in DTN, because in our case one cannot consider any of
the spin-spin interactions as weak. Also, the use of hard-core
bosons for the description of the behavior of DTN for $T > T_N(H)$
cannot help because from the viewpoint of their collective behavior
they can be regarded as fermions (i.e., only one fermion, or
hard-core boson can be in one state). For $T< T_N(H)$ in DTN, we
definitely cannot use hard-core bosons (e.g., for hard-core bosons
BEC is impossible, however, see Ref.[\onlinecite{Zapf}] for DTN).
The advantage of our fermionic description of DTN for $T >T_N(H)$
(and the mentioned bosonic description for $T < T_N(H)$), compared
to the use of only a bosonic description of low-energy spin
excitations, is that in our approach both fermions and bosons are
basically {\em non-interacting}. Hence, they have all features of
standard fermions and bosons. For strongly interacting bosons, which
is the case for hard-core bosons or the Holstein-Primakoff
representation for $T > T_N(H)$, one cannot, strictly speaking, use
directly the bosonic character of these excitations. We finally want
to note, that in our calculations we never used the symmetry of the
wave function, the other difference between fermions and bosons.

In summary, our magnetic and magneto-acoustic studies of the
quantum spin-chain magnet NiCl$_2$-4SC(NH$_2$)$_2$ show that the
behavior of the observed properties at $T > T_N(H)$ can be well
described by an effective 1D {\em fermionic} model of low-lying
spin excitations. This fact, together with previous results
showing the bosonic 3D behavior of the magnetic excitations in DTN for $T
< T_N(H)$, \cite{Zapf} suggests the presence of a
crossover from a fermionic to a bosonic character of the magnetic
excitations close to the quantum critical points. The fermionic
and bosonic nature of the magnetic excitations is related to the
short-range correlations in the spin chains and to the long-range
three-dimensional order, respectively.

We thank S.~A.~Zvyagin for stimulating discussions. A.A.Z.
acknowledges the support from the Ukrainian Fundamental Research
State Fund (F25.4/13).

\end{document}